\input psfig.tex
\documentstyle[a4,12pt,epsf]{article}

\def\a{{\alpha}}
\def\b{{\beta}}
\def\c{{\gamma}}
\def\d{{\delta}}
\def\s{{\sigma}}
\def\r{{\rho}}

\begin{document}
\baselineskip .80cm

\vbox{\vspace{6mm}}

\begin{center}{
{\large \bf
COVARIANT GENERALISATION OF CODAZZI-RAYCHAUDHURI AND AREA CHANGE
EQUATIONS FOR RELATIVISTIC BRANES}\\[7mm]
Elias Zafiris\\
{\it Theoretical Physics Group \\
Imperial College \\
The Blackett Laboratory \\
London SW7 2BZ \\
U.K. \\
e.mail:e.zafiris@ic.ac.uk \\}
\vspace{2mm}

}\end{center}
\vspace{4mm}
\begin{abstract}
In this paper we derive the generalisations of Gauss-Codazzi, Raychaudhuri and area
change equations for classical relativistic  branes and
multidimensional fluids in arbitrary background manifolds with
metricity and torsion. The kinematical description we develop is
fully covariant and based on the use of projection tensors tilted with
respect to the brane worldsheets.
\end{abstract}

\renewcommand{\baselinestretch}{2}

\section{Introduction}

A large variety of physical systems is possible to be modeled as
relativistic branes of an appropriate dimension propagating in a fixed
background manifold.
In general a $(p-1)$ brane is to be understood as a dynamical system
defined in terms of fields with support confined to a $p-dim$
worldsheet surface $S$ in a background spacetime manifold $M$ of dimension
$n\geq p$ [1,2,3].

Stachel's idea of matter of multidimensional objects
[4,5,6] is a generalisation of the notion of point particles
to matter, whose elementary constituents are extensive objects. In
turn a multidimensional fluid of extended objects on $M$ is defined by
a congruence  of $p-dim$ worldsheets of $(p-1)-dim$
branes.

A great deal of interest in brane models comes from Cosmology, and
especially from theories of structure formation in the early universe
[7,8]. Essentially the relativistic theory of classical branes may be
applied to vacuum defects produced by the Kibble mechanism [9], with
interesting cosmological and astrophysical implications [10].

In recent years there has been a significant amount of work regarding
the development of a kinematical description of deformations of
the worldsheet spanned in the background manifold by a relativistic
brane [11,12,13,14,15,16]. The main motivation for a proper
kinematical description originates from a clear analogy. To be
concrete, it is well established today that the proof of the existence
of spacetime singularities in General Relativity relies on the
consequences obtained from Raychaudhuri equations for geodesic
congruences [17,18,19]. In brane theories the notion of the point
particle and the associated with it worldline, gets replaced by the
notion of extensive objects with their corresponding
worldsheets. Thus, in principle, it would be possible to derive the
generalised Raychaudhuri equations for brane worldsheet congruences
and arrive at analogous singularity theorems in Classical relativistic
Brane theory.

The great majority of the earlier approaches, have been expressed in
a highly gauge dependent notation, as well as after the introduction
of special reference systems, involving specifically adapted
coordinates and frames, taylored to the embedding of particular
surfaces, that require the simultaneous use of many different kinds of
indices. On the other hand the advantages of a covariant kinematical
description have been emphasized by Carter. Moving in this direction,
Carter, based on traditional surface embedding theory
[20,21,22,23,24], has developed a kinematical formalism which is, however,
in his words, ``designed to be a balance
compromise between frame-dependent and tensorial
expressions'' [1,15,16,25,26,27]. Motivated by the significance of
developing a proper mathematical machinery for the study of brane
models, in this paper we construct a generalised fully covariant
kinematical framework for branes and multidimensional fluids in
manifolds with metricity and torsion, using projection tensor techniques.

Moreover all the previous approaches, impose major restrictions on the
kind of brane models considered. More concretely, on the first place, 
the background manifold has to be endowed with a metric
tensor which is at least invertible. On the second place, they assume that the background manifold connection has to be
metric compatible and torsion free, thus excluding any discussion of
brane models in the higher dimensional manifolds of unified field
theories, which use non metric compatible and non torsion free
connections. Finally, they assume that the projection onto the brane worldsheet is
normal, thus excluding the treatment of null brane worldsheets [28] where
the normality condition is not satisfied. In this paper we remove
these restrictive assumptions. One key idea is to define two
different projection gradients which become equal for normal
projections.

Using the generalised kinematical framework we construct, we manage to
obtain the generalisations of the Gauss-Codazzi and Raychaudhuri
equations, governing the behaviour of brane worldsheet congruences in
manifolds with metricity and torsion, as well as the law governing
their generalised area change. In this way we recover and greatly
generalise the results of the current literature, for example those of
Capovilla, Guven [13,14], and Carter [15,16], establishing an elegant
covariant kinematical formalism at the same time.

The stucture of the paper is as follows:

In section 2 we develop our covariant kinematical framework for brane
worldsheets congruences. In more detail, in subsection 2.1  we introduce the descriptive elements of a general brane
model and present the projected index formalism as well as the
geometrical objects notation  we are going to
use. In subsection 2.2 we derive the kinematical quantities associated
with the brane worldsheet projection tensor fields. In subsection 2.3
we construct the intrinsic and extrinsic brane worldsheet projected
 covariant derivatives. In subsection 2.4  we decompose the metricity
and 
torsion tensors  and derive the generalised Weingarden identity. In
subsection 2.5 we consider
the decomposition of the Riemannian curvature w.r.t. the brane
worldsheet and derive  the generalisations of the Gauss-Codazzi and Raychaudhuri
equations in covariant form. In section 3  we study the brane worldsheet decomposition of
the Lie derivatives of the various geometrical objects and we obtain
the law governing the generalized area change. Finally we summarise
and conclude in section 4.

\section{Covariant kinematical framework for brane worldsheet congruences}

\subsection{Preliminaries and Notation}

The basic descriptive element of a general $(p-1)$ brane model as
localised on a $p-dim$ worldsheet in an $n-dim$ curved background
manifold, is the $rank-p$ operator $Z_{\mu\nu}$ of projection on the
worldsheet. The projection tensor field $Z$ assigns to each point $P$
of the manifold, a map $Z(P):T_P \rightarrow T_P$, on the tangent
space $T_P$, such that
\begin{equation}
Z(P)^2=Z(P)
\end{equation}

If $I$ is the identity map, the tensor field
\begin{equation}
V=I-Z
\end{equation}

is also a projection tensor field which we call the complement of
$Z$. An immediate consequence of (1) and (2) is $ZV=VZ=0$.

So long as there is a regular metric tensor we can require the
projection tensor field $Z_{\mu\nu}$ to be normal or equivalently
\begin{equation}
Z_{\mu\nu}=Z_{\nu\mu}
\end{equation}

We note that the above condition cannot be imposed on the projection
tensor onto a null brane worldsheet.

In cases where the normality condition (3) can be imposed, like the
projection onto a timelike brane worldsheet, the $rank-p$ operator
$Z_{\mu\nu}$ of tangential projection onto the brane worldsheet,
represents the metric induced on the $p-dim$ worldsheet by its
embedding in the $n-dim$ background manifold, whereas the
complementary $rank -(n-p)$ operator $V_{\mu\nu}$ of projection
orthogonal to the worldsheet, represents the projected metric on the
$(n-p) - dim$ quotient space if we consider a spacetime filling
congruence of brane worldsheets.

Moreover the dimension of the projection tensor $Z$ at the point $P$
is defined to be the dimension of the subspace $ZT_P$ which is the
tangent space of the brane worldsheet. An immediate consequence of
this definition is
$$p={Z^\a}_{\a},\quad\quad n-p={V^\a}_{\a}$$
In turn, the projection tensor $Z$ onto a null brane worldsheet is
characterised by a projected metric tensor (or a projected inverse
metric tensor) which is noninvertible and thus fails to define a
metric on the subspace $ZT_P$.

Because $Z$ and $V$ have the same formal properties, it is possible to
arrange all the expressions so that they are preserved by an exchange
of $Z$ and $V$. We have arranged the notation so that the exchange of
a projection tensor by its complement can be carried out easily.

The projection of high-rank tensors related to the kinematical
framework of branes often have lengthy expressions involving many
index contractions. The expressions become much more elegant and
transparent if we adopt a compact projected index notation formalism.

Thus we follow the conventions:
Each tensor index that is meant to be contracted with an index on the
projection tensor $Z$ is marked by the symbol $\wedge$, while each
index which is meant to be contracted with an index on the projection
tensor $V$ is marked by the symbol $\vee$. For example
\begin{equation}
{X^{{\buildrel \wedge \over \gamma}{\buildrel \vee \over
\alpha}}}_{\buildrel \wedge \over \delta}:={Z^\c}_\r {V^\a}_\s
{Z^\b}_\d {X^{\r \s}}_\b
\end{equation}

We also notice the distinction between a tensor, which inhabits a
particular projection subspace which we call entirely projected, and a tensor which is the
result of projecting the corresponding background manifold tensor into
that subspace.
An entirely projected tensor carries a projection label for each of
its tensor indices. The projection labels stand for projection tensors
and indicate the projection identities which are associated with each
tensor index. We can, by convention, abbreviate and use a single
symbol $(\a)$ to stand for an index-label pair $\buildrel \wedge \over
\a$ or $\buildrel \vee \over \a$. We interpret the summation convention on a repeated abbreviated symbol
to imply a sum over both the visible index value $\a$ and the
invisible projection label, and denote it by $[\a]$. This notation has
the advantage that a whole collection of entirely projected tensors is
organised into a single geometrical object with appropriate labels
according to the above conventions. Moreover organising entirely
projected tensors into geometrical objects has the advantage of making
very compact expressions, which are identical  in structure to
familiar unprojected tensor expressions.

Let us agree that the symbol ${{\buildrel p \over A}^{(\r)}}_{(\mu)
(\nu)}$ denotes the projected background manifold ${A^\r}_{\mu \nu}$
tensor and that the symbol ${A^{(\r)}}_{(\mu)
(\nu)}$ denotes the entirely projected background manifold ${A^\r}_{\mu \nu}$
tensor. The projected $A$ tensor geometrical object can be considered
as a representation of the background manifold tensor $A$.  In what follows it will
become clear that whenever a background manifold tensor is defined in
terms of covariant derivatives, its corresponding entirely projected
geometrical object will be different from the geometrical object of
the projected background manifold tensor.

In the following we use the definitions and notation of the projected and entirely
projected geometrical objects to set a covariant kinematical framework
for branes and multidimensional fluids in manifolds having metricity
and torsion, i.e. dealing with the most general case of brane models
that can occur.

\subsection{Kinematical quantities associated with the brane worldsheet
projection tensor fields}

The first step to set up a covariant kinematical framework for
classical relativistic branes is to find a way to express the first
derivative of the projection tensor field on the $p-dim$ brane
worldsheet as well as of its complement, in terms of tensors that obey
projection identities.

We define the projection gradient to be the tensor:
\begin{equation}
{Z^\a}_\s {Z^\r}_\b \nabla_\r {Z^\s}_\c=\nabla_{\buildrel \wedge \over
\b} {Z^{\buildrel \wedge \over \a}}_\c:={{z_\c}^{\buildrel \wedge \over
\a}}_{\buildrel \wedge \over \b} 
\end{equation}

The definition of the projection gradient is projected explicitly on
two of its three indices. However we can easily show that it obeys the
projection identity:
\begin{equation}
\nabla_{\buildrel \wedge \over
\b} {Z^{\buildrel \wedge \over \a}}_\c=\nabla_{\buildrel \wedge \over
\b} {Z^{\buildrel \wedge \over \a}}_{\buildrel \vee  \over \c}
\end{equation}

Another way to project the covariant derivative of a projection
tensor yields the tensor: 
\begin{equation}
{Z^\s}_{\c} {Z^\r}_\b \nabla_\r {Z^\a}_\s=\nabla_{\buildrel \wedge \over
\b} {Z^\a}_{\buildrel \wedge \over \c}:={z^\a}_{{\buildrel \wedge
\over \c}{\buildrel \wedge \over \b}}
\end{equation}

When a metric is available and $Z$ is a normal projection tensor this projection gradient is exactly the same as the previous one, with the indices
appropriately raised and lowered.

In turn it obeys the projection identity
\begin{equation}
{z^\a}_{{\buildrel \wedge \over \c}{\buildrel \wedge \over
\b}}={z^{\buildrel \vee \over \a}}_{{\buildrel \wedge \over \c}{\buildrel \wedge \over
\b}}
\end{equation}

In addition to the projection gradient associated with the brane
worldsheet projection operator $Z$, the same definition yields 
projection gradients  associated with the complementary projection
tensor $V$.
\begin{equation}
\nabla_{\buildrel \vee \over
\b} {V^{ \a}}_{\buildrel \vee \over \c}:={v^\a}_{{\buildrel \vee \over \c}{\buildrel \vee \over
\b}}={v^{\buildrel \wedge \over \a}}_{{\buildrel \vee \over \c}{\buildrel \vee \over
\b}}
\end{equation}
\begin{equation}
\nabla_{\buildrel \vee \over
\b} {V^{\buildrel \vee \over \a}}_\c:={{v_\c}^{\buildrel \vee \over
\a}}_{\buildrel \vee \over \b} ={{v_{\buildrel \wedge \over
\c}}^{\buildrel \vee \over \a}}_{\buildrel \vee \over \b} 
\end{equation}

Now we can  use the decomposition of the identity tensor $I=Z+V$ to
force a decomposition of the projection gradient w.r.t. the brane worldsheet.
\begin{equation}
\nabla_\d{Z^\a}_{\c}=\nabla_{\buildrel \wedge \over \d} {Z^{\buildrel
\wedge \over \a}}_{\buildrel \vee \over \c} + \nabla_{\buildrel \wedge \over \d} {Z^{\buildrel
\vee \over \a}}_{\buildrel \wedge \over \c} -\nabla_{\buildrel \vee \over \d} {V^{\buildrel
\wedge \over \a}}_{\buildrel \wedge \over \c}- \nabla_{\buildrel \vee \over \d} {V^{\buildrel
\vee \over \a}}_{\buildrel \wedge \over \c}
\end{equation}

or equivalently
\begin{equation}
\nabla_\d{Z^\a}_{\c}={{z_{\buildrel \vee \over \c}}^{\buildrel \wedge \over \a}}_{\buildrel \wedge \over \d}+
{z^{\buildrel \vee \over \a}}_{{\buildrel \wedge \over \c}{\buildrel \wedge \over
\d}}-{v^{\buildrel \wedge \over \a}}_{{\buildrel \vee \over \c}{\buildrel \vee \over
\d}}-{{v_{\buildrel \wedge \over
\c}}^{\buildrel \vee \over \a}}_{\buildrel \vee \over \d} 
\end{equation}

The decomposition of the complementary projection gradient $\nabla V$
is given by the complement of this expression, namely by exchanging
$Z$ and $V$, as well as $\wedge$ and $\vee$ everywhere.

Because each projection gradient has two indices which project into
the same subspace, we can contract these two indices and thus extract
symmetric and antisymmetric parts. From them we can define the
kinematical quantities associated with a projection tensor field.

The antisymmetric part reads
\begin{equation}
\nabla_{[\buildrel \wedge \over \d} {Z^{\buildrel
\vee \over \a}}_{\buildrel \wedge \over \c]} :={ {\buildrel \wedge \over
\omega}_{\c \d}}^\a
\end{equation}

and we call it the projection vorticity tensor.

The symmetric part reads
\begin{equation}
\nabla_{(\buildrel \wedge \over \d} {Z^{\buildrel
\vee \over \a}}_{\buildrel \wedge \over \c)} :={{\buildrel \wedge \over
\theta}_{\c \d}}^\a
\end{equation}

and we call it the projection expansion rate tensor.

When a metric tensor is available for raising and lowering indices we can also define a trace part
\begin{equation}
\nabla_{(\buildrel \wedge \over \d} {Z^{\buildrel
\vee \over \d}}_{\buildrel \wedge \over \c)} :={ {\buildrel \wedge \over
\theta}_{\c}}
\end{equation}

which we call the projection divergence, as well as a trace-free
symmetric part
\begin{equation}
{{\buildrel \wedge \over \s}_{\c \d}}^\a :={{\buildrel \wedge \over
\theta}_{\c \d}}^\a -{1 \over p} Z_{\c \d} { {\buildrel \wedge \over
\theta}^{\a}}
\end{equation}

which we call the projection shear tensor.

\subsection{Intrinsic and extrinsic brane worldsheet projected covariant derivatives}

We consider a vector field $u$ which obeys the projection identity
$Zu=u$,  meaning that $u(P)$ $\in ZT_P$ for every point P in the
background  manifold, namely the tangent space of the brane
worldsheet. The part of the covariant derivative $\nabla u$ reflecting
the way in which the vector field $u$ changes within the tangent space
of the brane worldsheet is defined to be the brane worldsheet  projected covariant
derivative $Du$ with components
\begin{equation}
D_\d u^{\buildrel \wedge \over \a}={Z^\a}_\r \nabla_\d u^{\buildrel \wedge \over \r}
\end{equation}

On the other hand the part of the covariant derivative ignoring the
way in which $u$ changes within the tangent space of the brane
worldsheet is defined to be the brane worldsheet antiprojected covariant
derivative with components
\begin{equation}
{\tilde D}_\d u^{\buildrel \wedge \over \a}={V^\a}_\r \nabla_\d u^{\buildrel \wedge \over \r}
\end{equation}

We can easily show that the derivatives defined above give zero when
they act on the projection tensor fields themselves
\begin{equation}
D_\d{Z^\a}_\b=0 \quad\quad\quad D_\d{V^\a}_\b=0 
\end{equation}

It is to be noted that these derivatives act only on tensor fields
belonging to particular projected subspaces and thus obey the product
rule only for products of these tensors.

The projected and antiprojected covariant derivatives are not entirely
projected geometrical objects. The desired entirely  projected objects are
the following
\begin{equation}
D_{\buildrel \wedge \over \d} u^{\buildrel \wedge \over
\a}={Z^\r}_{\d} D_\r u^{\buildrel \wedge \over \a}
\end{equation}

which we call the intrinsic brane worldsheet projected covariant
derivative,
\begin{equation}
D_{\buildrel \vee \over \d} u^{\buildrel \wedge \over
\a}={V^\r}_{\d} D_\r u^{\buildrel \wedge \over \a}
\end{equation}

which we call the extrinsic brane worldsheet projected covariant
derivative,
\begin{equation}
{\tilde D}_{\buildrel \wedge \over \d} u^{\buildrel \wedge \over
\a}={Z^\r}_{\d} {\tilde D}_\r u^{\buildrel \wedge \over \a}
\end{equation}

which we call the intrinsic brane worldsheet antiprojected covariant
derivative, and finally
\begin{equation}
{\tilde D}_{\buildrel \vee \over \d} u^{\buildrel \wedge \over
\a}={V^\r}_{\d} {\tilde D}_\r u^{\buildrel \wedge \over \a}
\end{equation}

which we call the extrinsic brane worldsheet antiprojected covariant
 derivative.

Using the geometrical objects notation we find that  the full decomposition of the background manifold
connection w.r.t. the brane worldsheet into entirely projected parts is determined by:
\begin{equation}
{\buildrel p \over \nabla}_{(\d)}u^{(\a)}=D_{(\d)}u^{(\a)}+u^{[\s]} {\Xi^{(\a)}}_{[\s]
(\d)}
\end{equation}
where
\begin{equation}
{\Xi^{(\a)}}_{(\s)
(\d)}:={\nabla Z^{(\a)}}_{[\mu] (\d)} {Y^{[\mu]}}_{(\s)}
\end{equation}
and 
\begin{equation}
{Y^\s}_{\r}:={Z^\s}_{\r}-{V^\s}_{\r}
\end{equation}

In the above formula the geometrical object ${\Xi^{(\a)}}_{(\s)
(\d)}$ does not change sign under  complementation and stands as the
generator of correction terms in the relationship between the
covariant and the entirely projected w.r.t. the brane worldshet
derivatives. If we compute the components of ${\Xi^{(\a)}}_{(\s)
(\d)}$ we obtain

$$ {\Xi^{\buildrel \wedge \over \b}}_{{\buildrel \wedge \over
\a}{\buildrel \wedge \over \d}}=0, \quad \quad 
{\Xi^{\buildrel \wedge \over \b}}_{{\buildrel \vee \over
\a}{\buildrel \wedge \over \d}}=-{{z_{\buildrel \vee \over
\a}}^{\buildrel \wedge  \over \b}}_{\buildrel \wedge  \over \d}$$

$$ {\Xi^{\buildrel \vee \over \b}}_{{\buildrel \wedge \over
\a}{\buildrel \wedge \over \d}}={z^{\buildrel \vee \over \b}}_{{\buildrel \wedge \over \a}{\buildrel \wedge \over
\d}}, \quad \quad 
{\Xi^{\buildrel \vee \over \b}}_{{\buildrel \vee \over
\a}{\buildrel \wedge \over \d}}=0$$

$$ {\Xi^{\buildrel \wedge \over \b}}_{{\buildrel \wedge \over
\a}{\buildrel \vee \over \d}}=0, \quad \quad 
{\Xi^{\buildrel \wedge  \over \b}}_{{\buildrel \vee \over
\a}{\buildrel \vee \over \d}}={v^{\buildrel \wedge \over \b}}_{{\buildrel \vee \over \a}{\buildrel \vee \over
\d}}$$

$$ {\Xi^{\buildrel \vee \over \b}}_{{\buildrel \wedge \over
\a}{\buildrel \vee \over \d}}==-{{v_{\buildrel \wedge \over
\a}}^{\buildrel \vee  \over \b}}_{\buildrel \vee  \over \d}, \quad \quad 
{\Xi^{\buildrel \vee  \over \b}}_{{\buildrel \vee \over
\a}{\buildrel \vee \over \d}}=0$$

\subsection{Decomposition of the metricity and torsion w.r.t. the brane worldsheet}

We consider that the background  manifold is endowed with a
form-metric $g^{\mu \nu}$. The metricity tensor is defined by the
relation
\begin{equation}
{Q^{\mu \nu}}_{\r}=-\nabla_{\r} g^{\mu \nu}
\end{equation}

In turn we define the entirely projected metricity geometrical object
w.r.t. the brane worldsheet by the relation
\begin{equation}
{Q^{{(\mu)}{(\nu)}}}_ {(\r)}:=-D_{(\r)} {g^{(\mu) (\nu)}}
\end{equation}

In the above definition the entirely projected geometrical object
${Q^{{(\mu)}{(\nu)}}}_ {(\r)}$ includes the intrinsic metricity
${Q^{{\buildrel \wedge \over \mu}{\buildrel \wedge \over
\nu}}}_{\buildrel \wedge \over \r}$ associated with the subspace
$ZT_P$, the intrinsic metricity ${Q^{{\buildrel \vee \over
\mu}{\buildrel \vee \over
\nu}}}_{\buildrel \vee \over \r}$ associated with the subspace
$VT_P$, as well as the mixed projected metricities.

Using the geometrical objects notation we find that the decomposition of the full background manifold
metricity w.r.t. the brane worldsheet into entirely projected parts is
determined by the formula:
\begin{equation}
{{\buildrel p \over Q}^{{(\mu)}{(\nu)}}}_ {(\d)}
={Q^{{(\mu)}{(\nu)}}}_ {(\d)}-2g^{[\r] \big( (\mu)} {\Xi^{(\nu) \big)}}_{[\r] (\d)}
\end{equation}

Formula (29) is equivalent to the following system of equations in the
projected index formalism:
\begin{equation}
{{\buildrel p \over Q}^{{\buildrel \wedge \over \mu}{\buildrel \wedge \over
\nu}}}_{\buildrel \wedge \over \d}={Q^{{\buildrel \wedge \over
\mu}{\buildrel \wedge \over \nu}}}_{\buildrel \wedge \over \d}+ g^{{\buildrel \wedge \over
\mu}{\buildrel \vee \over \r}}{{z_{\buildrel \vee \over
\r}}^{\buildrel \wedge \over \nu}}_{\buildrel \wedge \over \d} +
g^{{\buildrel \vee \over
\r}{\buildrel \wedge \over \nu}}{{z_{\buildrel \vee \over
\r}}^{\buildrel \wedge \over \mu}}_{\buildrel \wedge \over \d} 
\end{equation}

\begin{equation}
{{\buildrel p \over Q}^{{\buildrel \wedge \over \mu}{\buildrel \wedge \over
\nu}}}_{\buildrel \vee \over \d}={Q^{{\buildrel \wedge \over
\mu}{\buildrel \wedge \over \nu}}}_{\buildrel \vee \over \d}- g^{{\buildrel \wedge \over
\mu}{\buildrel \vee \over \r}}{v^{\buildrel \wedge \over
\nu}}_{{\buildrel \vee \over \r}{\buildrel \vee \over \d}} -
g^{{\buildrel \vee \over
\r}{\buildrel \wedge \over \nu}}{v^{\buildrel \wedge \over
\mu}}_{{\buildrel \vee \over \r}{\buildrel \vee \over \d}} 
\end{equation}

\begin{equation}
{{\buildrel p \over Q}^{{\buildrel \wedge \over \mu}{\buildrel \vee \over
\nu}}}_{\buildrel \wedge \over \d}={Q^{{\buildrel \wedge \over
\mu}{\buildrel \vee \over \nu}}}_{\buildrel \wedge \over \d}- g^{{\buildrel \wedge \over
\mu}{\buildrel \wedge  \over \r}}{z^{\buildrel \vee \over
\nu}}_{{\buildrel \wedge \over \r}{\buildrel \wedge \over \d}}  +
g^{{\buildrel \vee \over
\r}{\buildrel \vee \over \nu}}{{z_{\buildrel \vee \over
\r}}^{\buildrel \wedge \over \mu}}_{\buildrel \wedge \over \d} 
\end{equation}

\begin{equation}
{{\buildrel p \over Q}^{{\buildrel \wedge \over \mu}{\buildrel \vee \over
\nu}}}_{\buildrel \vee \over \d}={Q^{{\buildrel \wedge \over
\mu}{\buildrel \vee \over \nu}}}_{\buildrel \wedge \over \d}+ g^{{\buildrel \wedge \over
\mu}{\buildrel \wedge \over \r}}{{v_{\buildrel \wedge \over
\r}}^{\buildrel \vee \over \nu}}_{\buildrel \vee \over \d} -
g^{{\buildrel \vee \over
\r}{\buildrel \vee \over \nu}}{v^{\buildrel \wedge \over
\mu}}_{{\buildrel \vee \over \r}{\buildrel \vee \over \d}} 
\end{equation}

and their complements.

We note that if the background manifold is endowed with a regular
metric tensor and the connection is metric compatible, the metricity
tensor vanishes, and thus the l.h.s. of the above equations equal
 zero. Moreover, for a normal projection tensor field the mixed
projected metrics vanish. Finally we observe that the intrinsic and
the mixed projected metricities do not necessarily vanish for a
non-normal projection tensor field even if the connection is metric compatible.

The torsion tensor is defined by the relation:
\begin{equation}
[\nabla_\nu, \nabla_\mu]\phi={S^\r}_{\mu \nu} \nabla_\r \phi
\end{equation}

The above definition can be decomposed into entirely projected parts
w.r.t. the brane worldsheet by defining the entirely projected torsion
tensors according to:
\begin{equation}
[D_{(\nu)}, D_{(\mu)}]\phi:={S^{(\r)}}_{(\mu) (\nu)} D_{(\r)} \phi
\end{equation}

Since $Z$ is the projection tensor onto the brane worldsheet the
quantity $ [D_{\buildrel \wedge \over \nu}, D_{\buildrel \wedge \over
\mu}] \phi $ is related to the torsion of the intrinsic geometry on
this surface.

In the definition (35) the entirely projected geometrical object $
{S^{(\r)}}_{(\mu) (\nu)} $ includes the intrinsic torsion
${S^{\buildrel \wedge \over \r}}_{{\buildrel \wedge \over
\mu}{\buildrel \wedge \over \nu}}$ associated with the subspace
$ZT_P$, the intrinsic torsion ${S^{\buildrel \vee \over \r}}_{{\buildrel \vee \over
\mu}{\buildrel \vee \over \nu}}$ associated with the subspace
$VT_P$, as well as the mixed projected torsions.

Using the geometrical objects notation we find that the decomposition of the full background manifold torsion w.r.t. the
brane worldsheet into entirely projected parts is determined by the
formula:
\begin{equation}
{{\buildrel p \over S}^{(\r)}}_{(\mu) (\nu)}={S^{(\r)}}_{(\mu)
(\nu)}-2 {\Xi^{(\r)}}_{\big[(\mu) (\nu) \big]}
\end{equation}

Formula (36) is equivalent to the following system of equations in the
projected index formalism:
\begin{equation}
{{\buildrel p \over S}^{\buildrel \wedge \over \r}_{{\buildrel \wedge \over \mu}{\buildrel
\wedge \over \nu}}}={S^{\buildrel \wedge \over \r}_{{\buildrel \wedge \over \mu}{\buildrel
\wedge \over \nu}}} 
\end{equation}
\begin{equation}
{{\buildrel p \over S}^{\buildrel \vee \over \r}_{{\buildrel \wedge \over \mu}{\buildrel
\wedge \over \nu}}}={  S^{\buildrel \vee \over
\r}_{{\buildrel \wedge \over \mu}{\buildrel \wedge \over \nu}}} -2{{\buildrel \wedge \over \omega}^\r}_{\mu \nu}  
\end{equation}
\begin{equation}
{{\buildrel p \over S}^{\buildrel \wedge  \over \r}_{{\buildrel \wedge \over \mu}{\buildrel
\vee \over \nu}}}={ S^{\buildrel \wedge \over
\r}_{{\buildrel \wedge \over \mu}{\buildrel \vee \over \nu}}}- {{z_{\buildrel \vee \over \nu}}^{\buildrel \wedge \over
\r}}_{\buildrel \wedge \over \mu}
\end{equation}

together with their complements.

Furthermore if there is a regular metric tensor, and the background manifold
connection is torsion free we obtain:
\begin{equation}
{{\buildrel p \over S}^{\buildrel \wedge \over \r}_{{\buildrel \wedge \over \mu}{\buildrel
\wedge \over \nu}}}={{\buildrel p \over S}^{\buildrel \vee \over \r}_{{\buildrel \wedge \over \mu}{\buildrel
\wedge \over \nu}}}={{\buildrel p \over S}^{\buildrel \vee \over \r}_{{\buildrel \wedge \over \mu}{\buildrel
\vee \over \nu}}}=0
\end{equation}

or equivalently
\begin{equation}
{  S^{\buildrel \wedge \over \r}_{{\buildrel \wedge \over \mu}{\buildrel
\wedge \over \nu}}}=0 
\end{equation}
\begin{equation}
{  S^{\buildrel \vee \over
\r}_{{\buildrel \wedge \over \mu}{\buildrel \wedge \over \nu}}} =2{{\buildrel \wedge \over \omega}^\r}_{\mu \nu}  
\end{equation}
\begin{equation}
{  S^{\buildrel \wedge \over
\r}_{{\buildrel \wedge \over \mu}{\buildrel \vee \over \nu}}}= {{z_{\buildrel \vee \over \nu}}^{\buildrel \wedge \over
\r}}_{\buildrel \wedge \over \mu}
\end{equation}

The $\wedge \wedge- $ projection of the torsion
definition reads:
\begin{equation}
[D_{\buildrel \wedge \over \nu},D_{\buildrel \wedge \over
\mu}]\phi=2{{\buildrel \wedge \over \omega}^\r}_{\mu \nu} D_{\buildrel
\vee  \over \r}\phi +{S^\r}_{{\buildrel \wedge \over \mu}{\buildrel
\wedge \over \nu}} \nabla_\r \phi
\end{equation}

In an adapted frame the $\wedge  \wedge-$ projection of the torsion definition
takes the form
\begin{equation}
{([e_n,e_m]-(2{\Gamma^r}_{[mn]}+{\Sigma^r}_{mn})e_r-{\Sigma^R}_{mn}e_R)}\phi=0
\end{equation}

The above expression provides the needed relation with the Frobenius
theorem. Concretely it shows that two vector fields with values in the
subspace $ZT_P$ have a commutator which lays in the same subspace if
and only if the mixed projected torsion tensor ${\Sigma^R}_{mn}$ is
zero. We call this condition generalised Weingarden identity. Then the Frobenius theorem guarantees that the subspace $ZT_P$
is tangent to a submanifold which is identified as the brane
worldsheet. As a consequence, if the background manifold connection is
torsion free,  the vanishing of the mixed projected
torsion tensor ${\Sigma^R}_{mn}$ implies, according to (42), that the vorticity tensor
${{\buildrel \wedge \over \omega}^\r}_{\mu \nu}$ should vanish.

\subsection{Decomposition of the Riemann curvature w.r.t. the brane worldsheet}

It is possible to relate the geometry of the projection tensor field
onto the brane worldsheet with the Riemannian geometry that it  inhabits,
by decomposing the spacetime Riemann tensor with respect to the
projection.

The curvature tensor is defined by the equation
\begin{equation}
u^\r {{R_\r}^\c}_{\a\b}=([\nabla_\b,\nabla_\a]-{S^\r}_{\a\b}
\nabla_\r) u^\c
\end{equation}

We can decompose the above definition, and form all of its independent
projections. In order to do this we first define the entirely
projected curvature geometrical object by the action of the entirely
projected covariant derivatives on the entirely projected w.r.t. the
brane worldsheet vector fields according to:
\begin{equation}
u^{[\r]} {{R_{[\r]}}^{(\c)}}_{(\a)(\b)}:=\big ([D_{(\b)},D_{(\a)}]-{S^{[\r]}}_{(\a)(\b)}
D_{[\r]} \big)u^{(\c)}
\end{equation}

In the above definition the first two pairs of indices of the entirely
projected curvature geometrical object must belong to the same
projected subspace. So we demand that 
$${{{R_{{\buildrel \wedge \over
\r}}}^{\buildrel \vee \over \c}}}_{(\a)(\b)}={{{R_{{\buildrel \vee \over
\r}}}^{\buildrel \wedge \over \c}}}_{(\a)(\b)}=0$$

Since the projection tensor $Z$ is surface forming it is clear that 
${{ R_{\buildrel \wedge \over \r}}^{\buildrel \wedge \over
\c}}_{{\buildrel \wedge \over \a}{\buildrel \wedge \over \b}}$ is the
intrinsic curvature tensor of the brane worldsheet. In an adapted
frame the components of ${{ R_{\buildrel \wedge \over \r}}^{\buildrel
\wedge \over \c}}_{{\buildrel \wedge \over \a}{\buildrel \wedge \over \b}}$ 
are given by the expression
\begin{eqnarray}
{{R_r}^c}_{ab}&=& e_b({\Gamma^c}_{ra})-e_a({\Gamma^c}_{rb})+{\Gamma^s}_{ra}
{\Gamma^c}_{sb}-{\Gamma^s}_{rb}
{\Gamma^c}_{sa} \cr
&-& {\Gamma^c}_{rs} ( 2{\Gamma^s}_{[ab]} -{\Sigma^s}_{ab})
-{\Sigma^S}_{ab} {\Gamma^c}_{ rS}
\end{eqnarray}

Similarly we call ${{ R_{\buildrel \vee \over \r}}^{\buildrel \vee \over
\c}}_{{\buildrel \vee \over \a}{\buildrel \vee \over \b}}$ the
intrinsic curvature associated with the projection tensor field
$V$, whereas we call  ${{ R_{\buildrel \wedge \over \r}}^{\buildrel \wedge \over
\c}}_{{\buildrel \wedge \over \a}{\buildrel \vee \over \b}}$  mixed projected curvature tensor.

Now it is possible to express the complete decomposition of the
background manifold Riemannian curvature tensor w.r.t. the brane worldsheet in
terms of the intrinsic and mixed projected curvature tensors. Using
the geometrical objects notation  the decomposition of the full background manifold
Riemann tensor w.r.t. the brane worldsheet into entirely projected
parts is determined by the formula:
\begin{eqnarray}
{{{\buildrel p \over R}_{(\r)}}^{(\c)}}_{(\a)(\b)}&=&{{
R_{(\r)}}^{(\c)}}_{(\a)(\b)}+2D_{ \big[ (\b)} {\Xi^{(\c)}}_{|(\r)|
(\a) \big]} - {S^{[\s]}}_{(\a) (\b)} {\Xi^{(\c)}}_{(\r)[\s]} \cr
&-&2
{\Xi^{(\c)}}_{[\d] \big[ (\a)} {\Xi^{[\d]}}_{|(\r)| (\b) \big]} 
\end{eqnarray}

Formula (49) is equivalent to the following system of equations in the
projected index formalism:

\begin{eqnarray}
{{{\buildrel p \over R}_{\buildrel \wedge \over \r}}^{\buildrel \wedge \over
\c}}_{{\buildrel \wedge \over \a}{\buildrel \wedge \over
\b}}={{ R_{\buildrel \wedge \over \r}}^{\buildrel \wedge \over
\c}}_{{\buildrel \wedge \over \a}{\buildrel \wedge \over \b}}
-{{z_{\buildrel \vee \over
\s}}^{\buildrel \wedge \over \c}}_{\buildrel \wedge \over \b}
{z^{\buildrel \vee \over
\s}}_{{\buildrel \wedge \over \r}{\buildrel \wedge \over \a}} 
+{{z_{\buildrel \vee \over
\s}}^{\buildrel \wedge \over \c}}_{\buildrel \wedge \over \a}
{z^{\buildrel \vee \over
\s}}_{{\buildrel \wedge \over \r}{\buildrel \wedge \over \b}} 
\end{eqnarray}

\begin{eqnarray}
{{{\buildrel p \over R}_{\buildrel \wedge \over \r}}^{\buildrel \wedge \over
\c}}_{{\buildrel \wedge \over \a}{\buildrel \vee \over
\b}}={{ R_{\buildrel \wedge \over \r}}^{\buildrel \wedge \over
\c}}_{{\buildrel \wedge \over \a}{\buildrel \vee \over \b}}
+{z^{\buildrel \vee \over
\s}}_{{\buildrel \wedge \over \r}{\buildrel \wedge \over \a}} 
{v^{\buildrel \wedge \over
\c}}_{{\buildrel \vee \over \s}{\buildrel \vee \over \b}} 
-{{z_{\buildrel \vee \over
\s}}^{\buildrel \wedge \over \c}}_{\buildrel \wedge \over \a}
{{v_{\buildrel \wedge \over
\r}}^{\buildrel \vee \over \s}}_{\buildrel \vee \over \b}
\end{eqnarray}

\begin{eqnarray}
{{{\buildrel p \over R}_{\buildrel \wedge \over \r}}^{\buildrel \wedge \over
\c}}_{{\buildrel \vee \over \a}{\buildrel \vee \over
\b}}={{ R_{\buildrel \wedge \over \r}}^{\buildrel \wedge \over
\c}}_{{\buildrel \vee \over \a}{\buildrel \vee \over \b}}
+{v^{\buildrel \wedge \over
\c}}_{{\buildrel \vee \over \s}{\buildrel \vee \over \a}} 
{{v_{\buildrel \wedge \over
\r}}^{\buildrel \vee \over \s}}_{\buildrel \vee \over \b}
-{v^{\buildrel \wedge \over
\c}}_{{\buildrel \vee \over \s}{\buildrel \vee \over \b}} 
{{v_{\buildrel \wedge \over
\r}}^{\buildrel \vee \over \s}}_{\buildrel \vee \over \a}
\end{eqnarray}

\begin{eqnarray}
{{{\buildrel p \over R}_{\buildrel \wedge \over \r}}^{\buildrel \vee \over
\c}}_{{\buildrel \wedge \over \a}{\buildrel \wedge \over \b}}=D_{\buildrel \wedge \over \b}{z^{\buildrel \vee \over
\c}}_{{\buildrel \wedge \over \r}{\buildrel \wedge \over \a}} -D_{\buildrel \wedge \over \a}{z^{\buildrel \vee \over
\c}}_{{\buildrel \wedge \over \r}{\buildrel \wedge \over \b}}
- {z^{\buildrel \vee \over
\c}}_{{\buildrel \wedge \over \r}{\buildrel \wedge \over \s}} {  S^{\buildrel \wedge \over \s}_{{\buildrel \wedge \over \a}{\buildrel
\wedge \over \b}}}+{{v_{\buildrel \wedge \over
\r}}^{\buildrel \vee \over \c}}_{\buildrel \vee \over \s}
 {S^{\buildrel \vee \over \s}_{{\buildrel \wedge \over \a}{\buildrel
\wedge \over \b}}}
\end{eqnarray}

\begin{eqnarray}
{{{\buildrel p \over R}_{\buildrel \wedge \over \r}}^{\buildrel \vee \over
\c}}_{{\buildrel \wedge \over \a}{\buildrel \vee \over \b}}=D_{\buildrel \vee \over \b}{z^{\buildrel \vee \over
\c}}_{{\buildrel \wedge \over \r}{\buildrel \wedge \over \a}} -D_{\buildrel \wedge \over \a}{{v_{\buildrel \wedge \over
\r}}^{\buildrel \vee \over \c}}_{\buildrel \vee \over \b}
 - {z^{\buildrel \vee \over
\c}}_{{\buildrel \wedge \over \r}{\buildrel \wedge \over \s}} { S^{\buildrel \wedge \over \s}_{{\buildrel \wedge \over \a}{\buildrel
\vee \over \b}}}+{{v_{\buildrel \wedge \over
\r}}^{\buildrel \vee \over \c}}_{\buildrel \vee \over \s}
{  S^{\buildrel \vee \over \s}_{{\buildrel \wedge \over \a}{\buildrel
\vee \over \b}}}
\end{eqnarray}

\begin{eqnarray}
{{{\buildrel p \over R}_{\buildrel \wedge \over \r}}^{\buildrel \vee \over
\c}}_{{\buildrel \vee \over \a}{\buildrel \vee \over \b}}=-D_{\buildrel \vee \over \b}{{v_{\buildrel \wedge \over
\r}}^{\buildrel \vee \over \c}}_{\buildrel \vee \over \a}+
D_{\buildrel \vee \over \a}{{v_{\buildrel \wedge \over
\r}}^{\buildrel \vee \over \c}}_{\buildrel \vee \over \b}
 - {z^{\buildrel \vee \over
\c}}_{{\buildrel \wedge \over \r}{\buildrel \wedge \over \s}} { S^{\buildrel \wedge \over \s}_{{\buildrel \vee \over \a}{\buildrel
\vee \over \b}}}+{{v_{\buildrel \wedge \over
\r}}^{\buildrel \vee \over \c}}_{\buildrel \vee \over \s}
{  S^{\buildrel \vee \over \s}_{{\buildrel \vee \over \a}{\buildrel
\vee \over \b}}}
\end{eqnarray}

and their complements.

We note that equations (50) and (53) are the generalisations of the
Gauss-Codazzi relations.  From equation (54) we can extract an
integrability condition in the following form:
\begin{equation}
{{R_{[{\buildrel \wedge \over \r}}}^{\buildrel \vee \over
\b}}_{{\buildrel \wedge \over \a}]{\buildrel \vee \over \b}}=
D_{\buildrel \vee \over \b} {{\buildrel \wedge \over \omega}^\b}_{\r
\a} +D_{[{\buildrel \wedge \over \a}}{{\buildrel \vee \over \theta}_{\r]}} 
 - {z^{\buildrel \vee \over
\b}}_{[{{\buildrel \wedge \over \r}|{\buildrel \wedge \over \s}}|} {{\buildrel r \over  S}^{\buildrel \wedge \over \s}_{{\buildrel \wedge \over \a}]{\buildrel
\vee \over \b}}}+{v_{[{\buildrel \wedge \over
\r}}}^{\buildrel \vee \over \c}_{|{\buildrel \vee \over \s}|}
{ S^{\buildrel \vee \over \s}_{{\buildrel \wedge \over \a}]{\buildrel
\vee \over \b}}}=0
\end{equation}

When the background manifold is endowed with a regular metric tensor,
the projection tensor is normal,  and the background manifold
connection is metric compatible and torsion free, we have only three
independent projections of the Riemannian tensor. The three independent projections together with symmetries of the
Riemann tensor as well as their complements provide a complete
decomposition of the Riemannian curvature.

Finally it is of interest  to consider the decomposition of the contracted
curvature, namely the Ricci tensor,
w.r.t. the brane worlsheet.

First we construct the $\wedge \wedge-$ projection of the Ricci curvature tensor.
\begin {eqnarray}
{\buildrel p \over R}_{{\buildrel \wedge \over \a}{\buildrel \wedge \over \b}}=
 R_{{\buildrel \wedge \over \a}{\buildrel \wedge \over \b}}
+D_{\buildrel \vee \over \s}{z^{\buildrel \vee \over
\s}}_{{\buildrel \wedge \over \a}{\buildrel \wedge \over \b}} 
+D_{\buildrel \wedge \over \b}{\buildrel \vee \over \theta}_\a \cr
+{z^{\buildrel \vee \over
\r}}_{{\buildrel \wedge \over \a}{\buildrel \wedge \over \s}} 
({{z_{\buildrel \vee \over
\r}}^{\buildrel \wedge \over \s}}_{\buildrel \wedge \over \b}-
{  S^{\buildrel \wedge \over \s}_{{\buildrel \wedge \over \b}{\buildrel
\vee \over \r}}})-{z^{\buildrel \vee \over
\s}}_{{\buildrel \wedge \over \a}{\buildrel \wedge \over \b}} 
{\buildrel \wedge \over \theta}_\s+
{{v_{\buildrel \wedge \over
\a}}^{\buildrel \vee \over \r}}_{\buildrel \vee \over \s}
{  S^{\buildrel \vee \over \s}_{{\buildrel \wedge \over \b}{\buildrel
\vee \over \r}}}
\end{eqnarray}

where $ R_{{\buildrel \wedge \over \a}{\buildrel \wedge \over \b}}$ is the intrinsic Ricci curvature
tensor associated with the brane worldsheet.

The $\vee \vee-$ projection of the Ricci curvature tensor is obtained by
taking the complements in (57), and corresponds to the intrinsic Ricci
curvature associated with the projection tensor field $V$.

By contracting ${\buildrel p \over R}_{{\buildrel \wedge \over \a}{\buildrel \wedge \over \b}}$ with the projected metric tensor 
$g^{{\buildrel \wedge \over \a}{\buildrel \wedge \over \b}}$
we obtain a generalisation of the Raychaudhuri equation in the form:
\begin{eqnarray}
g^{{\buildrel \wedge \over \a}{\buildrel \wedge \over \b}} 
{\buildrel p \over R}_{{\buildrel \wedge \over \a}{\buildrel \wedge \over \b}} &=&
{\buildrel \wedge \over R}+D_{\buildrel \vee \over \a} \theta^{\buildrel
\wedge \over \a}+D_{\buildrel \wedge \over \b} \theta^{\buildrel
\vee \over \b}-\theta_{\buildrel \wedge \over \s} \theta^{\buildrel
\wedge \over \s}+{z_{\buildrel \vee \over \s}}^{{\buildrel \wedge
\over \b}{\buildrel \wedge \over \r}} {z^{\buildrel \vee \over \s}}_{{\buildrel \wedge
\over \b}{\buildrel \wedge \over \r}} 
+{Q^{{\buildrel \wedge \over \a}{\buildrel \wedge \over \r}}}_{\buildrel \wedge \over \a} {\buildrel \vee \over
\theta}_{\r} \cr
&+& {Q^{{\buildrel \wedge \over
\a}{\buildrel \wedge \over \r}}}_{\buildrel \vee \over \b}  {z^{\buildrel \vee \over \b}}_{{\buildrel \wedge
\over \r}{\buildrel \wedge \over \a}} -{z^{{\buildrel \vee \over
\b}{\buildrel \wedge \over \a}}}_{\buildrel \wedge \over \s} 
{S^{\buildrel \wedge  \over \s}}_{{\buildrel \wedge
\over \a}{\buildrel \vee \over \b}}+ 
{v^{{\buildrel \wedge \over
\a}{\buildrel \vee \over \b}}}_{\buildrel \vee \over \s} 
{S^{\buildrel \vee  \over \s}}_{{\buildrel \wedge
\over \a}{\buildrel \vee \over \b}}
\end{eqnarray}

When the background manifold is endowed with a metric compatible and
torsion free connection as well as the projection tensor field is
normal, Raychaudhuri equation takes the form:
\begin{equation}
g^{{\buildrel \wedge \over \a}{\buildrel \wedge \over \b}} 
{\buildrel p \over R}_{{\buildrel \wedge \over \a}{\buildrel \wedge \over \b}} =
{\buildrel \wedge \over R}+D_{\buildrel \vee \over \a} \theta^{\buildrel
\wedge \over \a}+D_{\buildrel \wedge \over \b} \theta^{\buildrel
\vee \over \b}-\theta_{\buildrel \wedge \over \s} \theta^{\buildrel \wedge \over \s}-
{{v_{\buildrel \wedge \over
\a}}^{\buildrel \vee \over \s}}_{\buildrel \vee \over \b} {v^{{\buildrel \wedge \over
\a}{\buildrel \vee \over \b}}}_{{\buildrel \vee \over \s}} 
\end{equation}

In order to complete our analysis we mention the identities that the
entirely projected w.r.t. the brane worldsheet Torsion and Riemann curvature
geometrical objects obey. These are the following:

{\bf 1}: Entirely projected Torsion Bianchi identities:
\begin{equation}
{{R_{\big [(\r)}}^{(\c)}}_{(\a)(\b)\big ]}+D_{\big [(\r)}
{S^{(\c)}}_{(\a) (\b) \big ]} + {S^{(\c)}}_{[\s] \big [(\r)}
{S^{[\s]}}_{(\a) (\b) \big]}=0
\end{equation}

{\bf 2}: Entirely projected Curvature Bianchi identities:
\begin{equation}
D_{\big [(\r)} {{R_{|(\d)|}}^{(\c)}}_{(\a)(\b)\big ]} + 
{{R_{(\d)}}^{(\c)}}_{[\s] \big [(\r)} {S^{[\s]}}_{(\a) (\b) \big ]}=0 
\end{equation}

\section{Decomposition of Lie Derivatives w.r.t the brane worldsheet}

The Lie derivative of a vector field u with respect to a vector field
M, can  be obtained by the commutator
\begin{equation}
({\mathcal L}_M u)\phi=([M,u])\phi
\end{equation}

for every scalar field $\phi$ defined on the background  manifold.

The relationship between the Lie derivative and the covariant
derivative is provided by the formula
\begin{equation} 
({\mathcal L}_M u)^\d=M^\s \nabla_\s u^\d - u^\r ( \nabla_\r
M^\d-{S^\d}_{\r\s}M^\s)
\end{equation}

It is interesting to work out the brane worldsheet projection tensor
decomposition of the Lie derivative of the projection tensor $Z$.

{\it Proposition}: The full decomposition of the background manifold
Lie derivative of the projection tensor $Z$ w.r.t. the brane
worldsheet into entirely projected parts is detemined by the formula:
\begin{eqnarray}
{\mathcal L}_M {Z^{(\a)}}_{(\s)}&=& \frac{1}{2} \big[ {Y^{(\a)}}_{[\b]}
D_{(\s)} M^{[\b]} - {Y^{[\b]}}_{(\s)} D_{[\b]} M^{(\a)} \big] \cr
&+& \frac{1}{2} M^{[\d]} \big [{Y^{[\b]}}_{(\s)} {S^{(\a)}}_{[\b] [\d]}
-{Y^{(\a)}}_{[\b]} {S^{[\b]}}_{(\s) [\d]} \big]
\end{eqnarray}

{\it Proof}:

The relation between the Lie derivative of the projection tensor $Z$
with its covariant derivative is given by:
\begin{equation}
{\mathcal L}_M {Z^\a}_{\b}= M^\d \nabla_\d {Z^\a}_\b -{Z^\r}_\b
{\buildrel s \over  \nabla}_\r M^\a +{Z^\a}_\r {\buildrel s \over
\nabla}_\b M^\r
\end{equation}
where
\begin{equation}
{\buildrel s \over
\nabla}_\r M^\d={\nabla_\r} M^\d -{S^\d}_{\r \s} M^\s
\end{equation}

We consider the projection of the above relation w.r.t. the brane
worldsheet, using the geometrical objects notation to obtain:
\begin{equation}
{\buildrel p \over {\mathcal L}}_M {Z^{(\a)}}_{(\b)}=M^{[\d]} {\buildrel
p \over \nabla}_{[\d]} {Z^{(\a)}}_{(\b)} -{Z^{[\r]}}_{(\b)} {\buildrel
ps \over \nabla}_{[\r]} M^{(\a)} + {Z^{(\a)}}_{[\r]} {\buildrel ps \over
\nabla}_{(\b)} M^{[\r]}
\end{equation}

If we use the relation between the covariant derivative and the
entirely projected covariant derivative as well as the proposition
referring to the torsion tensor, we can calculate the r.h.s. of (67) in
terms of entirely projected geometrical objects as follows:
\begin{eqnarray}
{\buildrel p \over {\mathcal L}}_M {Z^{(\a)}}_{(\b)}&=&
 {Z^{(\a)}}_{[\r]} D_{(\b)} M^{[\r]}-{Z^{[\r]}}_{(\b)} D_{[\r]}
M^{(\a)} \cr
&+& M^{[\s]} \bigg [{Z^{[\r]}}_{(\b)} {S^{(\a)}}_{[\r] [\s]}-
{Z^{(\a)}}_{[\r]} {S^{[\r]}}_{(\b) [\s]} \bigg ]
\end{eqnarray}

Next we note that the following identity holds:
\begin{equation}
{\mathcal L}_M (Z+V)={\mathcal L}_M Z +{\mathcal L}_M V=0
\end{equation}

Equation (69) implies that (68) is antisymmetric under
complementation. This property becomes manifest if we make use of the
tensor ${Y^\a}_\b$ defined by (26), and thus write (68) equivalently in
the form:
\begin{eqnarray}
{\mathcal L}_M {Z^{(\a)}}_{(\s)}&=& \frac{1}{2} \bigg[ {Y^{(\a)}}_{[\b]}
D_{(\s)} M^{[\b]} - {Y^{[\b]}}_{(\s)} D_{[\b]} M^{(\a)} \bigg] \cr
&+& \frac{1}{2} M^{[\d]} \bigg [{Y^{[\b]}}_{(\s)} {S^{(\a)}}_{[\b] [\d]}
-{Y^{(\a)}}_{[\b]} {S^{[\b]}}_{(\s) [\d]} \bigg]
\end{eqnarray}
which completes the proof of the proposition.

If we assume that the background manifold connection is torsion free,
then the only non-zero projections of the Lie derivative of the
projection tensor $Z$ w.r.t. the brane worldsheet, in the projected
index notation, are given by:
\begin{equation} 
{{{({\mathcal L}_M Z})}^{\buildrel \wedge \over \a}}_{\buildrel \vee
\over \b}=-{(S_M)^{\buildrel \wedge \over \a}}_{\buildrel \vee
\over \b}+D_{\buildrel \vee \over \b} M^{\buildrel \wedge \over \a}+
{{z_{\buildrel \vee \over
\b}}^{\buildrel \wedge \over \a}}_{\buildrel \wedge \over \s}
M^{\buildrel \wedge \over \s}-2{{\buildrel \vee \over \omega}^\a}_{\b\s}
M^{\buildrel \vee \over \s}
\end{equation}

\begin{equation} 
{{{({\mathcal L}_M Z})}^{\buildrel \vee \over \a}}_{\buildrel \wedge
\over \b}={(S_M)^{\buildrel \vee \over \a}}_{\buildrel \wedge
\over \b}-D_{\buildrel \wedge \over \b} M^{\buildrel \vee \over \a}-
{{v_{\buildrel \wedge \over
\b}}^{\buildrel \vee \over \a}}_{\buildrel \vee \over \s}
M^{\buildrel \vee \over \s}+2{{\buildrel \wedge \over \omega}^\a}_{\b\s}
M^{\buildrel \wedge \over \s}
\end{equation}

where 
$${(S_M)^{\buildrel \star \over \a}}_{\buildrel \diamond
\over \b}:={S^{\buildrel \star \over \a}_{{\buildrel \diamond \over \b}{\buildrel
\wedge \over \s}}}M^{\buildrel \wedge \over \s}+
{S^{\buildrel \star \over \a}_{{\buildrel \diamond \over \b}{\buildrel
\vee \over \s}}}M^{\buildrel \vee \over \s}$$

We consider a vector field $u$ which obeys the projection identity
$Zu=u$. According to the definitions provided in the case of covariant
derivatives we define its brane worldsheet entirely  projected Lie
derivative by the formula
\begin{equation}
L_M u^{\buildrel \wedge \over \a}={Z^\a}_\r {\mathcal L}_M u^{\buildrel
\wedge \over \r} 
\end{equation}

whereas its brane worldsheet entirely antiprojected Lie derivative by the formula
\begin{equation}
{\tilde L}_Mu^{\buildrel \wedge \over \a}={V^\a}_\r {\mathcal L}_M u^{\buildrel
\wedge \over \r} 
\end{equation}

It is important to relate the brane worldsheet entirely tangentially projected Lie
derivative to the ordinary Lie derivative.

{\it Proposition}: The full decomposition of the background manifold Lie
derivative of an entirely projected vector field w.r.t. the brane
worldsheet into entirely projected parts is determined by:
\begin{equation}
({\buildrel p \over {\mathcal L}}_M u)^{(\a)}=L_M u^{(\a)}
-u^{[\r]} {{\Xi_M}^{(\a)}}_{[\r]}
\end{equation}
where
\begin{eqnarray}
{{\Xi_M}^{(\a)}}_{(\r)}&=& \frac{1}{2} \bigg[ {Y^{[\b]}}_{[\s]} D_{[\b]}
M^{(\a)} -{Y^{(\a)}}_{[\b]} D_{[\s]} M^{[\b]} \big] {Y^{[\s]}}_{(\r)} \cr
&+& \frac{1}{2} M^{[\c]} \bigg[ {Y^{(\a)}}_{[\b]} {S^{[\b]}}_{[\s] [\c]} -
{Y^{[\b]}}_{[\s]} {S^{(\a)}}_{[\b] [\c]} \bigg] {Y^{[\s]}}_{(\r)}
\end{eqnarray}

{\it Proof}:

We start with the expressions:
$$u^{(\a)}={Z^{(\a)}}_{[\r]} u^{[\r]}, \quad \quad if \quad u \in ZT_P$$
$$u^{(\a)}={V^{(\a)}}_{[\r]} u^{[\r]}, \quad \quad if \quad u \in VT_P$$

We Lie differentiate the above expressions
\begin{equation}
{\mathcal L}_M u^{(\a)}=({\mathcal L}_M {Z^{(\a)}}_{[\r]}) u^{[\r]} +
{Z^{(\a)}}_{[\r]}) {\mathcal L}_M u^{[\r]} 
\end{equation}
\begin{equation}
{\mathcal L}_M u^{(\a)}=-({\mathcal L}_M {Z^{(\a)}}_{[\r]}) u^{[\r]} +
{V^{(\a)}}_{[\r]}) {\mathcal L}_M u^{[\r]} 
\end{equation}
for $u \in ZT_P \quad or \quad VT_P$ respectively.

If we make use of the tensor ${Y^\a}_{\b}$ equations (77) and (78) can
be written in a single relation as follows:
\begin{equation}
{\mathcal L}_M u^{(\a)}=({\mathcal L}_M {Z^{(\a)}}_{[\s]})
{Y^{[\s]}}_{[\r]} u^{[\r]} +L_M u^{(\a)}
\end{equation}

where the following identity is satisfied
\begin{equation}
-({\mathcal L}_M {Z^{(\a)}}_{[\s]}){Y^{[\s]}}_{(\b)} = {Y^{(\a)}}_{[\s]}
({\mathcal L}_M {Z^{[\s]}}_{(\b)})
\end{equation}

By definition we have:
\begin{eqnarray}
{{\Xi_M}^{(\a)}}_{(\r)}&=& \frac{1}{2} \bigg[ {Y^{[\b]}}_{[\s]} D_{[\b]}
M^{(\a)} -{Y^{(\a)}}_{[\b]} D_{[\s]} M^{[\b]} \bigg] {Y^{[\s]}}_{(\r)} \cr
&+& \frac{1}{2} M^{[\c]} \bigg[ {Y^{(\a)}}_{[\b]} {S^{[\b]}}_{[\s] [\c]} -
{Y^{[\b]}}_{[\s]} {S^{(\a)}}_{[\b] [\c]} \bigg] {Y^{[\s]}}_{(\r)}
\end{eqnarray}

Hence if we combine (70), (79) and (81) we finally obtain:
\begin{equation}
({\buildrel p \over {\mathcal L}}_M u)^{(\a)}=L_M u^{(\a)}
-u^{[\r]} {{\Xi_M}^{(\a)}}_{[\r]}
\end{equation}

which completes the proof of the proposition.

Then
equation (82), after using (71) and (72) in the projected index notation
takes the form:

\begin{eqnarray}
{\mathcal L}_M u^{\buildrel \wedge \over \a}=
L_M u^{\buildrel \wedge \over \a}+(-{(S_M)^{\buildrel \wedge \over \a}}_{\buildrel \vee
\over \b}+D_{\buildrel \vee \over \b} M^{\buildrel \wedge \over \a}+
{{z_{\buildrel \vee \over
\b}}^{\buildrel \wedge \over \a}}_{\buildrel \wedge \over \s}
M^{\buildrel \wedge \over \s}-{{\buildrel \vee \over \omega}^\a}_{\b\s}
M^{\buildrel \vee \over \s} \cr
+{(S_M)^{\buildrel \vee \over \a}}_{\buildrel \wedge
\over \b}-D_{\buildrel \wedge \over \b} M^{\buildrel \vee \over \a}-
{{v_{\buildrel \wedge \over
\b}}^{\buildrel \vee \over \a}}_{\buildrel \vee \over \s}
M^{\buildrel \vee \over \s}+{{\buildrel \wedge \over \omega}^\a}_{\b\s}
M^{\buildrel \wedge \over \s})u^{\buildrel \wedge \over \b} 
\end{eqnarray}

Furthermore we can generalise the above notions and ask for the
decomposition of the Lie derivative of a general entirely projected
tensor w.r.t. the brane worldsheet. So let us consider the entirely
projected geometrical object ${X^{(\a) (\b)}}_{(\mu) (\nu)}$.
\begin{eqnarray}
{({\mathcal L}_M X)^{(\a) (\b)}}_{(\mu) (\nu)}&=& L_M {X^{(\a)
(\b)}}_{(\mu) (\nu)}-{X^{[\r] (\b)}}_{(\mu) (\nu)} {{\Xi_M}^{(\a)}}_{[\r]}
-{X^{(\a) [\r]}}_{(\mu) (\nu)} {{\Xi_M}^{(\b)}}_{[\r]} \cr
&+& {X^{(\a) (\b)}}_{[\r] (\nu)} {{\Xi_M}^{[\r]}}_{(\mu)} 
+{X^{(\a) (\b)}}_{(\mu) [\r]} {{\Xi_M}^{[\r]}}_{(\nu)} 
\end{eqnarray}

A significant application of the above decompositions  is obtained when we study  the evolution of
a timelike  brane worldsheet area element. Let's suppose that $dimZT_P=p$. The area
element on the brane worldsheet surface tangent to the subspace $ZT_P$
is the unit $p-$ form $Q$ in $Z^\star T_P$. We choose a vector field $N^a$
in $VT_P$ and a $p-$ form field field $S$ which is propagated along
the integral curves of $N^\a$ by Lie dragging so that it is a solution
of
$${\mathcal L}_N S=0$$

So long as the chosen Lie-dragged $p-$ form $S$ obeys

\begin{equation}
Z^* S \neq 0
\end{equation}

The $Z-$ area element $Q$ can be constucted from $S$. Since the
subspace$ZT_P$ is timelike we have
\begin{equation}
Q={(det[g_{{\buildrel \wedge \over \a}{\buildrel \wedge \over
\b}}])}^{1/2} Z^* S 
\end{equation}

where
\begin{equation}
(det[g_{{\buildrel \wedge \over \a}{\buildrel \wedge \over
\b}}]) \neq 0
\end{equation}
Because the entirely projected Lie derivative obeys 
\begin{equation}
L_NZ=0
\end{equation}
and the field $S$ has been defined by Lie dragging, we see that
\begin{equation}
L_N(Z^* S)=0
\end{equation}

Then the entirely projected Lie derivative of the tangent to the brane
worldsheet $Z-$ area element is 
\begin{equation}
L_N Q=-\frac{1}{2}g_{{\buildrel \wedge \over \a}{\buildrel \wedge \over
\b}} L_N g^{{\buildrel \wedge \over \a}{\buildrel \wedge \over
\b}} Q
\end{equation}

Furthermore we can calculate the quantity $L_N g^{{\buildrel \wedge
\over \a}{\buildrel \wedge \over \b}}$ as follows:
\begin{eqnarray}
{({\buildrel p \over {\mathcal L}}_N g)}^{(\a) (\b)}=L_N g^{(\a) (\b)}
-g^{[\r] (\b)} {{\Xi_N}^{(\a)}}_{[\r]}-g^{(\a) [\r]} {{\Xi_N}^{(\b)}}_{[\r]}
\end{eqnarray}

Hence we obtain:
\begin{eqnarray}
{({\buildrel p \over {\mathcal L}}_N g)}^{{\buildrel \wedge \over
\a}{\buildrel \wedge \over \b}}&=& L_N g^{{\buildrel \wedge \over
\a}{\buildrel \wedge \over \b}} -g^{{\buildrel \wedge \over
\r}{\buildrel \wedge \over \b}} {{\Xi_N}^{\buildrel \wedge \over
\a}}_{\buildrel \wedge \over \r} -g^{{\buildrel \wedge \over
\r}{\buildrel \wedge \over \b}} {{\Xi_N}^{\buildrel \wedge \over
\a}}_{\buildrel \vee  \over \r} \cr 
&-& g^{{\buildrel \wedge \over
\a}{\buildrel \wedge \over \r}} {{\Xi_N}^{\buildrel \wedge \over
\b}}_{\buildrel \wedge \over \r}-g^{{\buildrel \wedge \over
\a}{\buildrel \vee \over \r}}{{\Xi_N}^{\buildrel \wedge \over
\b}}_{\buildrel \vee  \over \r}  
\end{eqnarray}
where
$$ {{\Xi_N}^{\buildrel \wedge \over
\a}}_{\buildrel \wedge \over \b} =- {{S_N}^{\buildrel \wedge \over
\a}}_{\buildrel \wedge \over \b}$$
$${{\Xi_N}^{\buildrel \vee \over
\a}}_{\buildrel \wedge \over \b} =- {{S_N}^{\buildrel \vee \over
\a}}_{\buildrel \wedge \over \b}+
D_{\buildrel \wedge \over \b} N^{\buildrel \vee \over \a}+
{{v_{\buildrel \wedge \over
\b}}^{\buildrel \vee \over \a}}_{\buildrel \vee \over \s}
N^{\buildrel \vee \over \s}-2{{\buildrel \wedge \over \omega}^\a}_{\b\s}
M^{\buildrel \wedge \over \s}$$

and the complements of the above expressions.

The geometrical object 
${({\buildrel p \over {\mathcal L}}_N g)}^{(\a) (\b)}$  in terms of
the entirely projected covariant derivatives is given by:
\begin{eqnarray}
{({\buildrel p \over {\mathcal L}}_N g)}^{(\a) (\b)}&=&\big[ 2g^{[\r]
\big( (\mu)}
{S^{(\nu) \big)}}_{[\r] [\d]}  -{Q^{(\mu) (\nu)}}_{[\d]} \big ]
N^{[\d]} \cr
&-& g^{[\r] (\nu)} D_{[\r]} N^{(\mu)} -g^{(\mu) [\r]} D_{[\r]} N^{(\nu)} 
\end{eqnarray}

Now if we use equations (91), (92) and (93) we finally obtain:
\begin{eqnarray}
L_N g^{{\buildrel \wedge \over \a}{\buildrel \wedge \over \b}}&=&
-{Q_N}^{{\buildrel \wedge \over \a}{\buildrel \wedge \over
\b}}-2g^{{\buildrel \wedge \over \r}  {( \buildrel \wedge \over \a}}
D_{\buildrel \wedge \over \r} N^{\buildrel \wedge \over \b )}
+2g^{{\buildrel \vee  \over \r} {( \buildrel \wedge \over \b}}
{{z_{\buildrel \vee \over
\r}}^{\buildrel \wedge  \over \a) }}_{\buildrel \wedge  \over \s}
N^{\buildrel \vee \over \s} \cr
&+& 2 \big( g^{{\buildrel \vee  \over \r} {( \buildrel \wedge \over \a}}
{{z_{\buildrel \vee \over
\s}}^{\buildrel \wedge  \over \b})}_{\buildrel \wedge  \over \r} -
g^{{\buildrel \vee  \over \r} ( {\buildrel \wedge \over \a}}
{v^{\buildrel \wedge \over \b )}}_{{\buildrel \vee \over \r}{\buildrel
\vee \over \s}} \big) N^{\buildrel \vee \over \s}
\end{eqnarray}

where, the terms which vanish when we have torsionless and metric
compatible connections have been collected in the geometrical object
\begin{equation}
{{\buildrel p \over Q}_N}^{(\mu) (\nu)}=\big( {{\buildrel p \over
Q}^{(\mu) (\nu)}}_{[\d]} -g^{(\mu) [\r]} {{\buildrel p \over
S}^{(\nu)}}_{[\r] [\d]}-g^{[\r] (\nu)}{{\buildrel p \over
S}^{(\mu)}}_{[\r] [\d]} \big) N^{[\d]}
\end{equation}

Hence equation (90) gives the result:

\begin{equation}
g_{{\buildrel \wedge \over \a}{\buildrel \wedge \over
\b}} \big( g^{{\buildrel \wedge \over \r}({\buildrel \wedge
\over \a}}{{z_{\buildrel \vee \over \s}^{\buildrel \wedge \over \b)}}_{\buildrel \wedge \over \r}}-g^{{\buildrel \vee  \over \r} ( {\buildrel \wedge \over \a}}
{v^{\buildrel \wedge \over \b )}}_{{\buildrel \vee \over \r}{\buildrel
\vee \over \s}} \big) 
N^{\buildrel \wedge \over \s} Q=-L_N Q
\end{equation}

Finally by projecting the identity
\begin{equation}
g_{\a \b} g^{\r \a}= {\d^\r}_\b 
\end{equation} 

equation (96) takes the form:
\begin{equation}
N^\s {\buildrel \wedge \over \theta}_\s Q -\big( g_{{\buildrel \vee
\over a}{\buildrel \wedge \over \b}} g^{{\buildrel \wedge \over
\r}{\buildrel \vee \over \a}} {{z_{\buildrel \vee \over
\s}}^{\buildrel \wedge  \over \b}}_{\buildrel \wedge  \over \r}+g_{{\buildrel \wedge
\over a}{\buildrel \wedge \over \b}} g^{{\buildrel \vee \over
\r}{\buildrel \wedge  \over \a}} {v^{\buildrel \wedge \over \b )}}_{{\buildrel \vee \over \r}{\buildrel
\vee \over \s}} \big) N^\s Q
=-L_N Q
\end{equation}

Finally we note that if the projection tensor field onto the brane
worldsheet is normal the above equation is simplified as follows:
\begin{equation}
N^\s {\buildrel \wedge \over \theta}_\s Q =-L_N Q
\end{equation}

Equation (99) determines the relation between the divergence and the
rate of change of the timelike brane worldsheet  area element.

\section{Summary and Discussion}

In this paper we have constructed a fully covariant kinematical
framework for classical relativistic branes, satisfying needs arising
in the context of theories of topological defects structures, such as
cosmic strings, higher dimensional cosmic membranes, as well as
multidimensional fluids. All the previous approaches to such a
kinematical description had imposed major restrictions on the kind of
brane models considered, excluding any discussion of null branes, as
well as of branes in the higher dimensional manifolds of unified field
theories, which use non metric compatible and non torsion free
connections. Our treatment, based on the use of projection tensor
techniques, has removed all the restrictive assumptions. The basic
idea we have used, is the definition of two different projection
gradients which become equal for normal projections.

Our analysis has shown that when a projection tensor field is surface forming, the curvature
decomposition includes the generalisation of Gauss-Codazzi equations. When the
projection tensor field is not surface forming, multidimensional fluid
flow for example, the gradient of the projection tensor turns out to
be composed of such well-known quantities as the  vorticity  and the
expansion of fluid flow and the curvature decomposition leads to
the generalisation of Raychaudhuri equation. In the latter case
our framework generalises the kinematical notions of relativistic
Cosmology used in the context of General Relativity [29], to the case
of multidimensional fluids. In this way we have modeled the geometry
of arbitrary brane worldsheet congruences in manifolds with metricity
and torsion. In the context of our kinematical framework we have
managed to obtain the generalisations of the Gauss-Codazzi and
Raychaudhuri equations, as well as the law governing their generalised
area change in a covariant form.

\vspace{5.5cm}
\noindent{\bf Acknowledgements}

\bigskip

It is my great pleasure to thank  J.J. Halliwell for his
comments. This work has been partially supported by A.S. Onassis
public benefit foundation.
v

\end{document}